\documentclass[acmsmall]{acmart}

\AtBeginDocument{%
  \providecommand\BibTeX{{%
    \normalfont B\kern-0.5em{\scshape i\kern-0.25em b}\kern-0.8em\TeX}}}

\setcopyright{acmlicensed}
\acmDOI{10.1145/3698128}
\acmYear{2024}
\acmJournal{PACMHCI}
\acmVolume{8}
\acmNumber{ISS}
\acmArticle{528}
\acmMonth{12}
\received{2024-02-22}
\received[accepted]{2024-05-30}

\usepackage{verbatim}
\usepackage{_macros}

\showrevisions{BLUE}

\begin{document}

\title[Planar or Spatial: Exploring Design Aspects and Challenges for Presentations in VR with No-coding Interface]{Planar or Spatial: Exploring Design Aspects and Challenges for Presentations in Virtual Reality with No-coding Interface}


\author{Liwei Wu}
\orcid{0009-0009-5750-0691}
\affiliation{%
  \institution{University of Waterloo}
  \city{Waterloo}
  \country{Canada}
}
\email{a92wu@uwaterloo.ca}

\author{Yilin Zhang}
\orcid{0009-0002-0675-1545}
\affiliation{%
  \institution{University of Waterloo}
  \city{Waterloo}
  \country{Canada}
}
\email{yilinjz@cmu.edu}

\author{Justin Leung}
\orcid{0000-0001-5363-6609}
\affiliation{%
  \institution{University of Waterloo}
  \city{Waterloo}
  \country{Canada}
}
\email{justin.leung1@uwaterloo.ca}

\author{Jingyi Gao}
\orcid{0009-0002-2461-7251}
\affiliation{%
  \institution{University of Waterloo}
  \city{Waterloo}
  \country{Canada}
}
\email{aoggoa12138@gmail.com}

\author{April Li}
\orcid{0009-0008-4110-3600}
\affiliation{%
  \institution{University of Waterloo}
  \city{Waterloo}
  \country{Canada}
}
\email{aaprilli.sy@gmail.com}

\author{Jian Zhao}
\orcid{0000-0001-5008-4319}
\affiliation{%
  \institution{University of Waterloo}
  \city{Waterloo}
  \country{Canada}
}
\email{jianzhao@uwaterloo.ca}

\renewcommand{\shortauthors}{Liwei Wu, Yiling Zhang, Justin Leung, Jingyi Gao, April Li, and Jian Zhao}


\begin{abstract}
    The proliferation of virtual reality (VR) has led to its increasing adoption as an immersive medium for delivering presentations, distinct from other VR experiences like games and 360-degree videos by sharing information in richly interactive environments. However, creating engaging VR presentations remains a challenging and time-consuming task for users, hindering the full realization of VR presentation's capabilities. This research aims to explore the potential of VR presentation, analyze users' opinions, and investigate these via providing a user-friendly no-coding authoring tool. Through an examination of popular presentation software and interviews with seven professionals, we identified five design aspects and four design challenges for VR presentations. Based on the findings, we developed VRStory, a prototype for presentation authoring without coding to explore the design aspects and strategies for addressing the challenges. VRStory offers a variety of predefined and customizable VR elements, as well as modules for layout design, navigation control, and asset generation. A user study was then conducted with 12 participants to investigate their opinions and authoring experience with VRStory. Our results demonstrated that, while acknowledging the advantages of immersive and spatial features in VR, users often have a consistent mental model for traditional 2D presentations and may still prefer planar and static formats in VR for better accessibility and efficient communication. We finally shared our learned design considerations for future development of VR presentation tools, emphasizing the importance of balancing of promoting immersive features and ensuring accessibility.

\end{abstract}




\begin{teaserfigure}
    \centering
    \includegraphics[width=1\textwidth]{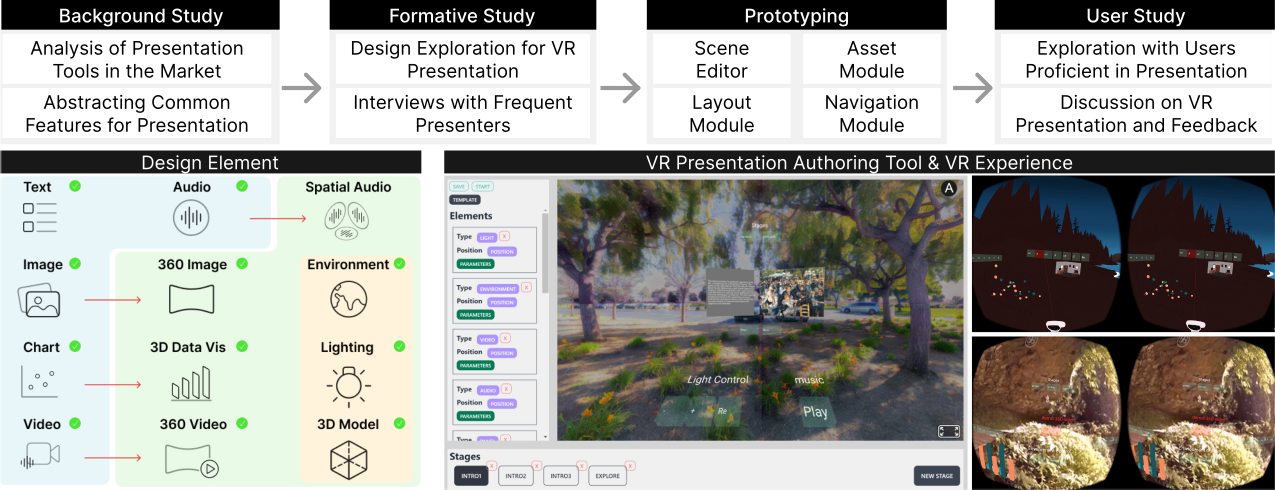}\vspace{-1ex}
    \caption{This research is conducted through four stages: (1) background study to identify common design aspects for VR presentations (\autoref{sec:background-study}); (2) interviews with professionals who frequently deliver presentations to investigate the design challenges and user perspectives for VR presentations (\autoref{sec:formative-study}); (3) prototyping an authoring tool to enable non-technical users to make VR presentations with ease (\autoref{sec:system}); (4) user studies with 12 participants (\autoref{sec:user-study}) to validate the findings in the formative study and collect users' perspectives on VR presentation. 
    }
    \label{fig:flow}
\end{teaserfigure}

\maketitle

\section{Introduction}
The significance of effective presentation cannot be overstated in various fields such as education and business, where information delivery and communication are crucial. 
While traditional 2D presentations have been well-established, there is growing interest in exploring the immersive potential of virtual reality (VR) for more engaging presentations \cite{Fuhrmann_2001, Peixoto_2021, mahalil_virtual_2019, tran_subjective_2017}.
With VR devices gaining widespread popularity, delivering presentations in VR environments is starting to draw interest from VR practitioners across platforms such as Microsoft Mesh \cite{microsoft_nodate}, Meta Horizon \cite{meta_nodate}, and Mozilla Hubs \cite{hubs_nodate}. 
Distinct from other popular VR experiences like gaming and 360-degree videos, delivering presentations in VR aims to leverage immersive and interactive VR features to enhance information communication and audience engagement, which we will refer to as \emph{VR presentation} in this paper. 
It can be also expected that an increasing number of non-technical users will start to explore and incorporate VR presentations into their work, ideally with no-coding user interfaces. 
However, most users have yet to explore the potential of VR presentations, leaving it a lack of opportunities to fully understand the common design aspects and challenges specific to users when engaging with VR presentations.

Moreover, during the authoring process, creating captivating VR presentations presents a unique set of challenges compared to traditional 2D presentations. 
For 2D presentations, presenters need to seamlessly integrate texts with high-quality media like images and videos \cite{zheng_telling_2022, yang_tactile_2020, he_comparing_2000, jokela_mobile_2008}, while adjusting layouts and navigation to suit their domain requirements and narration preferences \cite{thanyadit_tutor_2023, liu_igscript_2021, dow_presence_2007, boreczky_interactive_2000}.
Transitioning to VR introduces new elements like 3D objects and spatial interactions that significantly increase the conceptual and authoring complexity. 
While several platforms offer authoring tools designed for creating immersive VR experiences, these may be overly sophisticated for VR presentations and still pose challenges to manage and manipulate, especially for non-technical users.
This demands a deliberated design of VR presentation authoring tools with no-coding or low-coding interactions.
Further, many current VR presentations appear as enhanced 2D presentations in a large space, displaying 2D screens for images and videos. 
It is unclear whether this is due to a lack of appropriate authoring tools or simply user preferences, calling for further investigation into users' attitudes and perspectives on VR presentations. 

To fill the above gaps, in this work, we focus on exploring the potential of VR presentation around the following three research questions through four stages (\autoref{fig:flow}):
\begin{enumerate}
    \item[] \textbf{RQ1 - Design Aspects}: What are the common design elements for VR presentations? What are unique to VR and what can be transferred and extended from 2D presentations?
    \item[] \textbf{RQ2 - Challenges}: What are the concerns and challenges preventing users from authoring and engaging with VR presentations? 
    \item[] \textbf{RQ3 - Opinions}: What are users' perspectives on VR presentations? What strengths and constraints do they perceive about VR presentations?  
\end{enumerate}
To answer these RQs, in Stage 1 of the research, we conducted a background study on popular presentation tools to identify the key design aspects for VR presentations (RQ1).
In Stage 2, we interviewed seven professionals from three domains including university educators, designers, and business people who frequently deliver presentations, aiming to identify the design challenges as well as gain insights into their opinions regarding VR presentations (RQ2 \& 3).
Based on the interview findings, in Stage 3, we developed VRStory, a prototype authoring tool, which streamlines the process of creating interactive VR presentations without coding required. 
VRStory serves as a valuable research facilitator, investigating various strategies to address the identified challenges and empowering non-technical users to easily explore different design aspects of VR presentations tailored to their specific needs.
In Stage 4, we conducted a user study with VRStory, involving 12 participants who are proficient in presentation software, which aimed to validate the findings in the formative study and further collect the users' perspectives on VR presentations through a hands-on authoring session (RQ2 \& 3).
Finally, based on our findings, we also provide design considerations for future design and development for VR presentations.

\section{Related Work}
Our work bridges and adds to three main streams of research: 1) presentation tools for communication and storytelling, 2) authoring tools for VR experience specifically targeting non-technical users, and 3) content generation for creative and professional purposes. 

\subsection{Overview of Presentation Tools}
Over the last two decades, advancements in popular presentation tools have been significant \cite{roels_conceptual_2019}, moving from traditional tools such as Microsoft PowerPoint \cite{microsoft_ppt} to more interactive products like Prezi \cite{prezi} and Gather \cite{gather}. 
This evolution reflects a trend in which the characteristics of presentation tools are shifting from content-oriented to user-experience-oriented. This is evidenced by two factors: 1) the boundary between presentation and communication tools is becoming less distinct, and 2) the presentation style is moving from linear to non-linear \cite{NextSlidePlease_2012, HyperSlides_2013, MultiPresenter_2008}.

The main goal of a successful presentation is to effectively communicate information to its audience \cite{james_tips_2010}. However, due to the nature of presentations, the presenter often has more control over the process than the audience, which can create difficulties for the audience to understand the information \cite{reuss_powerpoint_2008}. 
To address this issue, some popular presentation tools in the market (\eg, Gather) have incorporated communication methods between the presenter and the audience, establishing an acknowledgment loop that helps maintain a balance between the two parties.

Recent studies have highlighted a growing trend toward presentation tools that enable greater speaker-audience interaction.
Presenters are increasingly integrating activities such as role-playing \cite{Gong_2021} and real-time Q\&A system \cite{Triglianos_2017} to facilitate immediate feedback from the audience and sustain their attention during presentations.
In addition, as Maleshkova et al. pointed out, 
interactive communication and idea exchange are crucial for some presentation types (\eg, art appreciation) \cite{Maleshkova_2013}.
Another trend in the last decade is a clear shift away from linear navigation (\eg, Powerpoint and Keynote) to Zoomable User Interfaces (ZUIs) \cite{lichtschlag_fly_2009, CounterPoint_2002, iMapping_2010}. 
ZUIs allow for greater user flexibility in selecting their own navigation order \cite{Roels_2013}. 
Additionally, the parallel structure of ZUIs enables users to better visualize the connections between topics \cite{Cheng_2015, J._2022, Li_2022, zdanovic_influence_2022}.
Also, recent studies have extended the shift: from a network/tree structure towards an immersive structure for presentations. 
This approach involves creating a space with presentation content, enabling users to obtain information spatially. 
By integrating AR/VR technologies, presenters can also create 3D environments that utilize multimedia in addition to images and text \cite{Speiginer_2019}, which further enhances the visualization capabilities of presentations \cite{roels_interactive_2017, roels_interactive_2023}.

Our work aims to continue exploring the potential of user-experience oriented presentations, as VR provides users with the flexibility of control over the presentation and offers various multimedia options for an immersive experience. 
Inspired by the previous studies, one of our goals is to explore the potential of VR presentations, particularly for non-technical users and identify the design features for VR presentations.

\subsection{Virtual Reality Authoring Tools}
While current VR devices still have limitations that hinder long-term use, research has signaled the potential of VR as an alternative working platform to traditional 2D displays \cite{biener_quantifying_2022}. 
Biener et al.'s study demonstrated the potential of VR for spatially arranging and manipulating information as the volume of 2D screens with interactions across touchscreens and VR HMDs \cite{biener_breaking_2020}, highlighting the promise of VR as a work tool due to its private and large display space.
However, an immersive VR experience goes beyond simply placing 2D screens in a virtual space and requires rich interactions and various forms of multimedia.
Creating engaging and fully-interactive VR experiences can be difficult for most users, especially those who lack VR development expertise.
While many current tools now offer easy-to-use functions for users to create static VR environments with platform-provided or self-imported 3D models like Mozilla Spoke, building fully dynamic behaviors and interactions still remains challenging \cite{artizzu_defining_2022}.
As Nebeling and Speicher pointed out, the massive tool landscape and the requirement for utilizing multiple tools for authoring VR experiences are the main challenges for non-technical users \cite{nebeling_trouble_2018}.
It is also known that the limitations of low-fidelity tools and the entry hurdles of high-fidelity can be common pitfalls even for professional VR designers and developers \cite{kraus_elements_2022}, not to mention non-technical users. 
It appears that users face a dilemma when choosing VR authoring tools: while low-fidelity tools like Spatial.io can create simple VR experiences without fully exploiting VR capabilities; high-fidelity tools like Unity enable fully interactive experiences but require coding that could be challenging to master and might be an overkill for users' particular domain tasks.

A systematic review by Coelho et al. analyzed 29 authoring tools for VR experiences \cite{coelho_authoring_2022}, revealing that many recent studies focused on enabling a broader audience to easily create, prototype, and develop interactive VR experiences. 
However, the authors emphasized the need for further research in specific application domains and underscored the importance of expert evaluation in these areas.
Various authoring techniques including physical prototyping \cite{nebeling_360proto_2019, speicher_designers_2021}, immersive authoring \cite{zhang_flowmatic_2020, zhu_save_2016,  xia_spacetime_2018}, and live or asynchronous collaborations \cite{thoravi_kumaravel_loki_2019, thoravi_kumaravel_transceivr_2020, nebeling_xrdirector_2020} have been explored in the literature.
Though these techniques empower the users to realize their ideas, it may be hard or take a long time for them to explore if they do not have clear ideas initially. 
Modular templates and functions, offered through building blocks \cite{takala_ruis_2014} or visual scripting \cite{zhang_flowmatic_2020, chen_entanglevr_2021}, may assist non-technical users in creating desired interactions and provide a starting point.
Particularly, the notion of End-User Development \cite{lieberman_end-user_2006} allows users to configure a peculiar VR experience to their needs by configuring VR experience templates created by experts \cite{artizzu_defining_2022}.
Immersive analytics, as a specialized area of study, primarily deals with examining and presenting data within VR environments. 
For readers looking to delve into this topic further, we suggest consulting Ens et al.'s extensive survey that outlines the challenges prevalent in immersive analytics and showcases current state-of-the-art toolkits \cite{ens_grand_2021}.

Inspired by previous works, our prototype authoring tool aims to strike a balance between creative freedom and efficiency by investigating modular building blocks and expert-created templates for customization.
This approach allows non-technical users to both easily create VR presentations with streamlined processes, while retaining the power to explore various VR features.

\subsection{Content Generation for Creative and Professional Purposes}
VR experiences require substantial multimedia resources, such as images and audio, to deliver high-quality and immersive results. 
This often necessitates professional skills and considerable effort, making it a significant barrier for non-technical users.
Fortunately, recent advancements in machine learning (ML) have shown promise in overcoming this hurdle by harnessing the generative capabilities of ML algorithms to produce outcomes that can even rival those achieved by human professionals.
For example, Zhou \etal\, utilized ML to generate new gaming experiences by reusing data from a digital game's dungeon \cite{Zhou_2021}. 
For images, multiple generative models in the industry \cite{dalle, stability_ai} and academia \cite{Badrinarayanan_2017, Kato_2019, Sagawa_2018} have enhanced the quality of images obtained through text-to-image and image-to-image generations. 
As for text generation, ML facilitates the process of generating personalized and easy-to-understand texts and dialogues \cite{HUANG_2022, Yang_2020}. 
In addition, ML can also be used in enhancing audio quality \cite{Li_2021} and generating music \cite{Raphael_2010, Mejtoft_2021}, as well as auto-transcribing videos for visually impaired users \cite{Yuksel_2020}. 

As the use of AI/ML in creative and professional contexts is still relatively new, there is limited literature exploring the utilization of ML-generated content for VR.
Enlightened by prior research, we envision that AI-assisted content generation could significantly enhance the ability of non-technical users to create and engage with VR presentations more effectively. 
To explore this potential and investigate users' perspectives, we incorporate content generation techniques for various media including speech, 2D and panorama images into our prototype. 

\section{Background Study} \label{sec:background-study}

\begin{figure*}[tb!]
    \centering
    \includegraphics[width=1\textwidth]{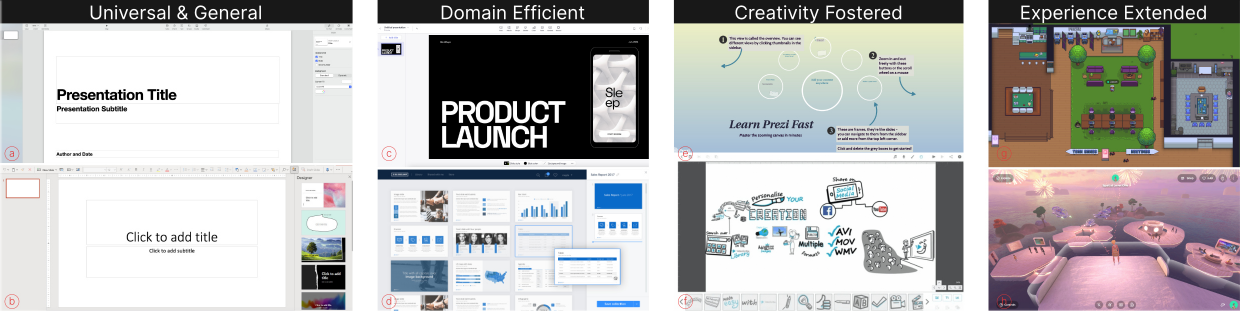}\vspace{-1ex}
    \caption{Examples of the presentation tools in four classified groups in our study: (a) Keynote, (b) PowerPoint, (c) Pitch, (d) SlideCamp, (e) Prezi, (f) VideoScribe, (g) Gather, and (h) Spatial.  }
    \label{fig:types}
\end{figure*}

As Stage 1 of our exploration (Figure~\ref{fig:flow}), we conducted a background study regarding common design aspects for presentations in two steps.
We first analyzed popular presentation tools and platforms in the market to identify the current trends and major design elements. 
Based on the analysis, we then derived common design aspects for VR presentations. 

\subsection{Analysis of Presentation Tools}
Being more exploratory than systematic at this stage, the whole research team began by individually collecting and reviewing a variety of presentation and communication tools. 
We used different sources of information, such as YouTube videos and online articles, to learn about the strengths and weaknesses of each tool. 
We then discussed our analysis as a team and evaluated each tool based on its target users, usage scenarios, and unique interactive elements.
This process resulted in a selection of 18 popular and representative presentation tools that covered a wide range of presentation styles and scenarios.
We classified these tools into four groups (Figure \ref{fig:types}): 
\begin{itemize}
    \item \textbf{Universal and Generalized} (G1): Tools like PowerPoint \cite{microsoft_ppt} and Keynote \cite{keynote} offer a wide range of features for creating effective presentations, but may demand a lot of work from users to adapt them to specific needs and domains.
    \item \textbf{Domain Efficient} (G2): Tools offer features that enhance basic functions and address users' specific needs for presentations in various scenarios, such as delivering a pitch with Pitch \cite{pitch}, maintaining consistent branding styles with Slidecamp \cite{slidecamp}, and choosing from a range of professional templates with Canva \cite{canva}.
    \item \textbf{Creativity Fostered} (G3): Tools enable more creative presentations than traditional 2D slides, such as controlling non-linear zoom-in-out style navigation with Prezi \cite{prezi}, creating presentation video with Videoscribe \cite{videoscribe} and visualizing data with infographics with PiktoChart \cite{piktochart}.
    \item \textbf{Experience Extended} (G4): Tools like Gather \cite{gather} and Spatial \cite{spatial}, unlike traditional 2D presentation tools, offer features that foster engagement and interaction among the audience in social and immersive settings.
\end{itemize}

\subsection{Design Aspects}

\begin{figure*}[tb!]
    \centering
    \includegraphics[width=1\textwidth]{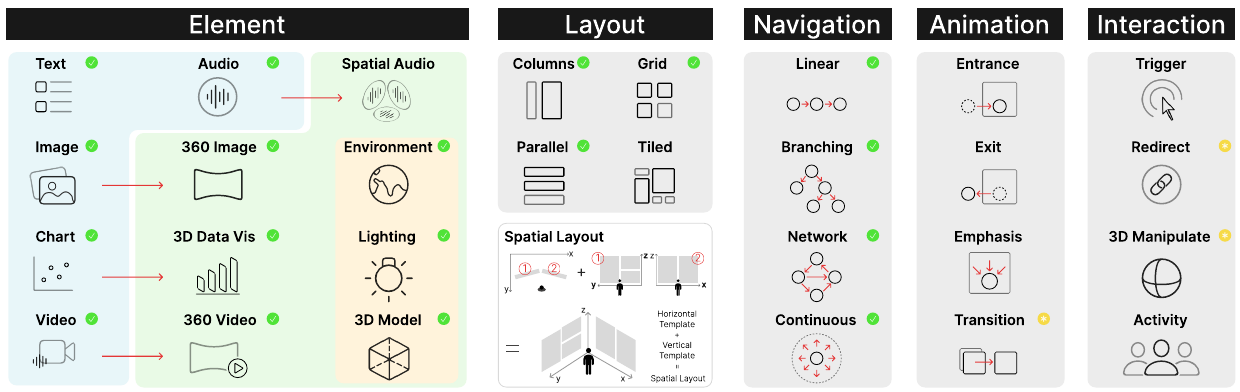}\vspace{-1ex}
    \caption{The background study identified five design aspects for VR presentation: 1) Elements serve as visual aids to convey information, including commonly used elements in traditional 2D presentations (blue area), elements used in VR presentations extending these 2D elements (green area), and unique elements specific to VR presentations (yellow area).
    2) Layouts describe the geometry relationships among the elements in one slide, including four base layouts that can be composed to suit different purposes of the slides, and spatial layout that can be composed by horizontal and vertical layouts.
    3) Navigation rules the methods one slide can move to other slides. 
    4) Animations and 5) Interactions are complementary aspects that can enhance the engagement and experience of presentations.
    Design Aspects that are supported by VRStory with customization are marked with green marks and predefined features offered by VRStory are marked with yellow marks.
    }
    \label{fig:features}
\end{figure*}

Through an analysis of common design elements from these collected presentation tools, we identified the five design aspects for effective presentations (Figure \ref{fig:features}).
Three aspects are considered by us as crucial aspects for an effective presentation: 
\begin{itemize}
    \item \textbf{Elements} such as text and images form the core building blocks of presentations, which are widely supported in all investigated tools. 
    Additionally, multimedia like audios and videos are also widely supported with default settings. More advanced elements, such as charts for data visualization, can be found in generalized tools like PowerPoint.
    \item \textbf{Layouts} exhibit the rules that define how multiple elements are graphically arranged on one single presentation slide.
    Most presentation tools in G1 and G2 offer layout templates that are designed to cater to the purpose of the slide by composing basic layouts like columns and grid, such as summary, agenda, or detail slides. Some tools in G3 have layout templates that are more tailored to the final delivery form, such as video (Powtoon) and long info diagram (Genially).
    \item \textbf{Navigation} reflects the patterns specifying how the slides are connected and how one slide can move to other slides.
    Most presentation tools in G1 and G2 encourage a linear navigation system where each slide can only go to the previous or next slide. Other presentation tools in G3 and G4 offer more diverse and free controls like network-like navigation (Prezi) and continuous game-like navigation (Gather). 
\end{itemize}
Two optional design aspects can improve audience engagement and presentation experience, though they are not necessary to include by default: 
\begin{itemize}
    \item \textbf{Animations} can be applied to individual elements or entire scenes to add motion or other effects.
    Tools in G1 and G3 groups tend to offer more basic animations and allow users to create more complex and delicate animations through the composition of basic animations. 
    On the other hand, tools in G2 tend to offer more opinion-based animations that are good to use without fine-tuning but allow for less customization.
    \item \textbf{Interaction} enables the audience to communicate with the speaker and engage with the presentation using different input methods (\eg, mouse clicks and hand gestures). 
    Many tools have limited default interactions, such as triggering element animations or redirecting to different slides through hyperlinks. Tools in G2 and G4 tend to offer more interaction tools to better support domain requirements (\eg, a speech timer in Pitch) or provide more engaging communication (\eg, editable sticky notes in Spatio) and activities (\eg, polls and quizzes in Prezi).
\end{itemize}

\subsection{Proof-of-Concept Prototype}
\begin{figure*}[tb!]
    \centering
    \includegraphics[width=1\textwidth]{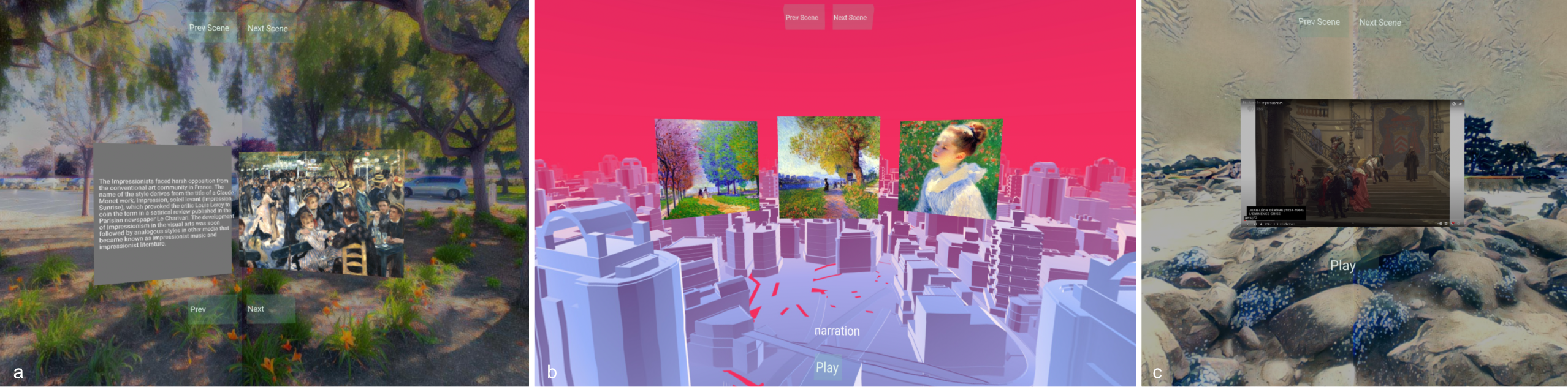}\vspace{-1ex}
    \caption{The prototype has three scenes with navigation buttons and two style transfer backgrounds (a, c). Design elements in each scene: (a) panels with switchable texts and images; (b) three text-generated images and a playable narration; and (c) a video with a play button. }
    \label{fig:prototype}
\end{figure*}

To summarize our background study, we created a proof-of-concept VR presentation prototype around the theme of Impressionism (\autoref{fig:prototype}) with A-Frame, a trending library for building WebXR experience.
We tested creating both 2D elements (\ie, text, images, videos) and 3D elements that are unique to VR (\ie, panorama images, lighting, spatial layout) in this prototype. 
We also explored ML methods such as image generation, style transfer, text abstraction, and voice generation to generate and process content for VR. 
The resulting presentation consisted of three VR scenes with panels displaying contents in three environments. 
Viewers were able to engage with the content by using their controllers to click buttons and transition between scenes.
The goal of building this proof-of-concept prototype was to facilitate our formative study (detailed in the next section) for learning more about users' perspectives on VR presentations. 

\section{Formative Study} \label{sec:formative-study}
In Stage 2 of our research (Figure~\ref{fig:flow}), to understand users' opinions towards VR presentations, we conducted a formative study with three groups of professionals who frequently deliver presentations in their daily jobs, including university educators, designers, and business people.

\subsection{Participants and Procedure}
Seven participants, comprising one female and six male professionals with a minimum of three years of professional work experience (\autoref{tab:participants}), were recruited through the network of a local research university.
The participants were selected based on three criteria: 1) they are knowledgeable about VR and ML technologies; 2) they have experienced VR during their professional work; and 3) they are interested in exploring new technologies in their domain.
Our goal was to better understand the prevalent practices and workflows for creating 2D presentations, and the presenters' perspectives on using VR presentations to meet their professional needs.

We conducted semi-structured interviews with our participants, where each session contained three parts. 
The first part focused on participants' experiences with creating professional presentations, aiming to gain insights into their current workflows and requirements.
The second part involved presenting the proof-of-concept VR presentation (Section 4.3) for feedback. 
Following that, we explained our production process and content generation methods, seeking participants' opinions on whether these techniques could be applicable to their work and identifying any missing features in the prototype.
The last part was an open discussion on VR presentation for professional scenarios to capture participants' concerns and visions.
Each interview lasted around one hour and each participant received \$20 for their time.

\begin{table*}[tb]
\footnotesize
    \centering
    \caption{Seven professionals from three different fields share their perspectives on VR presentations.}
    \begin{tabular}{p{0.02\linewidth} p{0.32\linewidth} p{0.58\linewidth}}
    \toprule
        \textbf{P\#} & \textbf{Professional Role} & \textbf{VR Experience} \\ 
    \midrule
    
        \textbf{P1} & Professor in science of information &
        active in VR research and has several published work about VR applications in education \\
        
         \textbf{P2} & Professor in film, television, and media  &
        one interactive VR film project where students can explore different camera techniques \\ 

        \textbf{P3} & Professor in architecture & 
        one interactive VR project where students can learn construction materials and methods\\ 

        \textbf{P4} & Junior UX designer in one tech company &  
        several professional projects involving interface and interaction design for the company's XR products\\ 
        
        \textbf{P5} & Junior architect in one international design agency & 
        several professional projects involving interactive architectural visualization in VR\\ 

        \textbf{P6} & Associate in investment banking and IPO service & 
        hands-on experience with various VR equipment and VR product demos\\ 

        \textbf{P7} & Senior business strategist and startup advisor & 
        hands-on experience with various VR equipment and VR product demos\\ 
         
     \bottomrule
    \end{tabular}
    
    \label{tab:participants}
\end{table*}

\subsection{Design Challenges} \label{sec:design-challenges}
Through a thematic analysis on the interview transcriptions, we identified four design challenges related to creating and utilizing VR presentations. 
The primary researcher first analyzed the transcriptions to build initial coding dimensions and generate initial codes. Then the primary researcher discussed the results with other team members around the research questions, refining the coding dimensions and the codes until reaching consensus.

\subsubsection{C1: Mental Model Challenges} \label{sec: mental1}
Non-technical participants may have limited familiarity with VR, which could hinder their ability to fully leverage the technology's capabilities when creating and engaging with VR presentations.
P4, P6, and P7, who frequently used presentations for business communication and collaboration, had an interest in VR presentations for enhanced 2D elements like larger 2D screens (P4) to exhibit content with greater detail and more space for visualizations (P6, P7) when presenting business data.
They preferred simple, linear layouts akin to traditional 2D presentations with one or a limited number of elements per slide, aimed at focusing audience attention on content. 
This suggested that these participants viewed VR presentations more as an upgrade from 2D without fully exploiting VR's unique capabilities.
On the other hand, P1, P2, P3 and P5, who are more experienced in 3D and VR, had more interests on leveraging the benefits of VR; they highlighted the need for more immersive and engaging elements like 3D models and 360-degree videos. 
As P1 commented, \qt{It is a little better than 2D just looking at slides in VR, but I still think it's not fully leveraging the capacities of VR.}
They also advocated for visually rich compositions and enhanced spatial navigation, inviting audiences to delve into presentations at their preferred pace and actively explore the content.

\subsubsection{C2: Accessibility Challenges} \label{sec: accessibility1}
While all the participants foresaw the potential active role of VR presentation in their professions, some of them (P2, P3, P5, and P6) highlighted the limited accessibility of VR presentation due to insufficient VR device availability and compatibility in workplaces.
P3 elaborated on this by mentioning that in his VR class, although sufficient equipment was provided, 80\% of students opted for desktops over VR headsets while learning, as they found it more efficient and user-friendly.
From another perspective, P2, who has a physical disability, offered their unique perspective \qt{I prefer not to be moving around or moving my arms. I prefer to have a more cinematic experience where I'm stationary and it's a dynamic interface but I don't have to do a lot of physical movement.} 

\subsubsection{C3: Technology Challenges}
Presently, developing interactive VR experiences heavily relies on game engines like Unity or programming 3D libraries like Three.js, which requires specialized technical skills and present substantial challenges to non-technical users. 
Except for P1, an experienced XR researcher, others strongly expressed the desire for coding-free authoring tools that enable increased independence, better control over final deliverables, and improved workflow efficiency. 
While P1 advocated the necessity of advanced customization through scripting, they acknowledged that non-technical users would benefit most from utilizing coding-free options.

\subsubsection{C4: Content Challenges} \label{sec: ai-concern1}
Creating an effective VR presentation involves high-quality media contents (e.g., images and videos) that align with the topic and style. 
Finding and customizing these resources can be challenging due to time constraints and the need for specialized software like Adobe Photoshop.
All participants acknowledged this difficulty in gathering suitable assets and there was a strong interest among participants in ML methods for content generation.
P1 envisioned an evolution where ML methods could generate entire VR presentations, commenting \qt{So now you have an ML tool. That is interesting. The ceiling is potentially endless, which is like maybe one day I can generate a whole VR environment for my teaching.}
P2 also emphasized the benefits of providing alternative content consumption methods through voice generation: \qt{I can see that (teaching over audio contents) as a possibility for students for reasons of access that would rather not interact with text but want the information in another format.}
Though content generation was thought to hold potentials, the need to maintain control over the content's accuracy, legality, and ethical standards was also emphasized by all the participants. 
P2, P3, P4, and P6 had reservations about using machine-generated images due to concerns regarding their visual quality and preferred relying on their own sources to meet their professional requirements.
P2, in particular, opposed the use of generated images for their film theory and history class, considering it ethically problematic, as stated \qt{I think it's ethically problematic. Faking general things that look like real, I don't like that idea. We have enough historical images available without having to create fabricated ones.}

\section{VRStory: A Prototype for VR Presentation Authoring} \label{sec:system}
Based on the background and formative study, in Stage 3 (Figure~\ref{fig:flow}), to explore the various design aspects and potential strategies for addressing the design challenges for VR presentations, we built VRStory, an authoring tool to empower users to create interactive VR presentations without any coding requirement.

\subsection{Designed Features}
VRStory empowers non-technical users to create VR presentations through desktop UI by utilizing customizable predefined templates and elements.
In traditional presentations, a \textit{Slide} typically represents two-dimensional content. 
To differentiate from this concept, we refer to each slide or individual scene within a VR presentation as a \textit{Stage} and the entire VR presentation as a \textit{Story} comprised of multiple stages.
To address the identified design challenges, VRStory is built with four interconnected modules, addressing the aforementioned four challenges: 
\begin{itemize}
    \item \textbf{Stage Editing} (C1, C3) : Modify and integrate predefined presentation elements in VR scenes through a user-friendly desktop interface without coding requirements, enabling users to easily discover and experiment with various 2D, 3D and VR features in their presentations.
    \item \textbf{Interactive Layout} (C2, C3) : Arrange content elements within VR scenes using customizable layout templates, ensuring an optimal viewing experience for stationary viewers with minimal physical movement.
    \item \textbf{Navigation Design} (C2, C3) : Prioritizes accessibility in VR presentations by enabling stationary viewing experience. Users can explicitly define navigation routes among stages for more spatial and dynamic presentation experience.
    \item \textbf{ML-Assisted Asset Management} (C4) : Streamline management of user-uploaded media assets such as images and 3D models, as well as create new assets including 2D and panorama images and audios with ML methods.
\end{itemize}

Further, VRStory offers \textbf{a rich set of elements} that are easy to discover and explore, designed to meet various VR presentation needs. These elements are accompanied by preconfigured interactions, minimizing the need for manual interaction creation.
VRStory provides elements in three categories (Figure \ref{fig:features}, Element):
1) \emph{2D elements} represent familiar elements from 2D presentations, presented as 2D planes in VR, such as text panels, images, audios, and videos; 
2) \emph{3D elements} extend 2D elements with the third dimension, including 360 images, 360 videos, and 3D data visualization;
3) \emph{Spatial elements} are exclusive to VR presentations and include the surrounding environment, ambient lighting, and 3D models, contributing to a more immersive and engaging VR experience.
VRStory simplifies VR element placement via its \textbf{spatial layout module} (Figure \ref{fig:features}, Spatial Layout). This feature caters to users of all levels by enabling simple composition with horizontal and vertical templates for beginners, while offering precise transformation parameter control for experts.  
VRStory also integrates \textbf{various navigation methods} to suit diverse user preferences, beginning with stationary viewing and progressing to dynamic locomotion options.
It provides a default linear progression through stages but offers customizable navigation options between any two stages, which gives users the ability to choose from basic linear navigation to intricate network navigation or any option in between.



\subsection{Features Walkthrough} \label{sec:system-walkthrough}
In this section, we walk through the main features of VRStory with a typical usage scenario. 
Let's assume Alex, a university student in art, wants to create a VR presentation for the topic ``Introduction to Impressionism.''

\begin{figure*}[tb]
    \centering
    \includegraphics[width=1\textwidth]{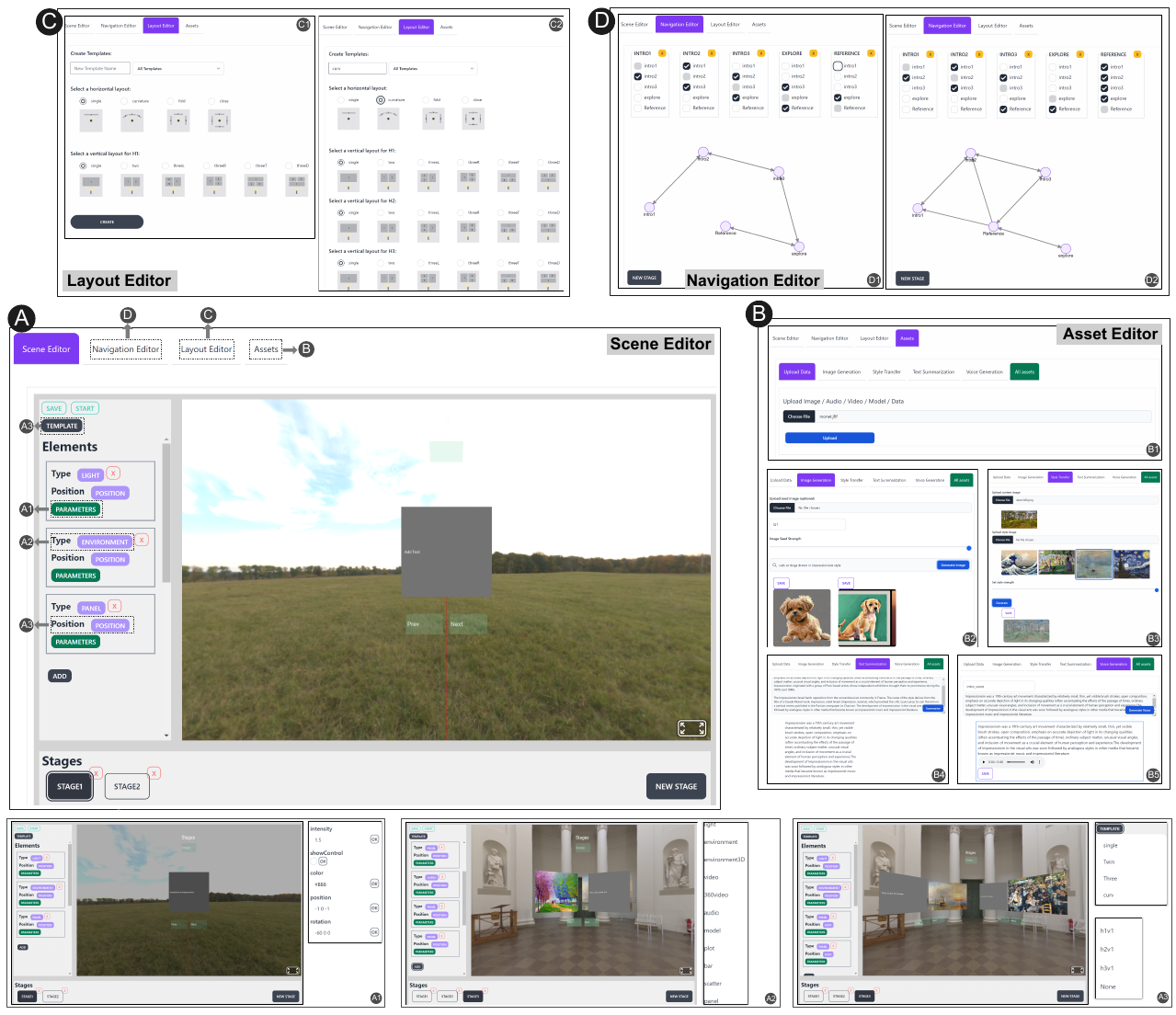}\vspace{-1ex}
    \caption{
    VRStory consists of four editors, namely: (A) the Scene Editor, which enables users to add and customize stage elements and utilize layout templates and media assets; (B) the Asset Editor, which facilitates the upload of users' own media assets and allows the creation of new images, texts, and audios using ML methods; (C) the Layout Editor, which generates layout placeholders for consumption in the Scene Editor, enabling the easy layout of stage elements in different positions; and (D) the Navigation Editor, which provides users with an intuitive interface for designing and visualizing navigation controls across all stages.
    }
    \label{fig:walkthrough}
\end{figure*}

\textbf{Initiate Story with Default Template and Stage.}
Alex begins crafting a VR story in VRStory's Scene Editor, starting with a template (Figure \ref{fig:walkthrough}, A). 
The template comprises two stages, each offering basic elements for building a VR presentation: environmental lighting, a central panel with text and an image and a panorama image background.
Alex promptly customizes the first stage by adjusting scene element parameters on the panels, changing the environmental lighting intensity to 1 for a darker scene and editing the panel text to ``Introduction to Impressionism'' (Figure \ref{fig:walkthrough}, A1).


\textbf{Create and Manage Media Assets.}
To enhance his presentation, Alex wants to add more images in the stage through the Asset Editor (Figure \ref{fig:walkthrough}, B). 
Here, he imports collected Monet paintings (Figure \ref{fig:walkthrough}, B1) and wants to show impressionism applied to everyday objects like pets. 
Due to limited online sources of impressionistic pet images, he uses the AI-powered content generation tools including text-to-image and style transfer (Figure \ref{fig:walkthrough}, B2 \& B3).
Realizing that Wikipedia text can be too lengthy, Alex utilizes the text abstraction tool to create summaries and voice generation tool to generate automated background narration using the summarized text (Figure \ref{fig:walkthrough}, B4 \& B5). 
Alex then switches between editors to create more content.

\textbf{Create New Stage and New Elements.}
Now Alex is satisfied with the first two stages and moves to create a new stage modifying an existing VRStory template.
Alex clicks ``add'' to insert a new element that is defaulted to a text panel. He can then change it to other types of elements by clicking ``type'' property in the scene editor (Figure \ref{fig:walkthrough}, A2). 

\textbf{Place Scene Elements through Layout Templates.}
Alex now uses the Layout Editor (Figure \ref{fig:walkthrough}, C) to generate layout placeholders using horizontal and vertical layout templates. 
Here Alex creates a layout template named ``Curv'' by selecting the ``curvature'' horizontal template containing three placeholders; for each horizontal placeholder, he chooses the ``single'' vertical template, which results in three placeholders in total (Figure \ref{fig:walkthrough}, C1 \& C2). 
Then, he applies this layout template in the Scene Editor and the previously added elements in the stage are automatically positioned accordingly (Figure \ref{fig:walkthrough}, A3).

\textbf{Design Explicit Navigation Control.}
Finally, Alex utilizes the Navigation Editor (Figure \ref{fig:walkthrough}, D) to define destinations for each stage through a connection graph, explicitly adding a single-direction link to the selected stage and managing the overall navigation flow (Figure \ref{fig:walkthrough}, D1 \& D2).

\subsection{Implementation}
VRStory is a web-based application comprising both a frontend client and a backend server.
The frontend consists of two components: a 3D VR scene renderer capable of supporting both 2D displays and VR headsets, and an editor used for customizing the scene. 
The 3D and VR features are implemented using A-Frame, a popular WebXR framework that leverages HTML segments to build VR scenes accessible on both desktops and VR headsets.
The data visualization elements are supported by BabiaXR, a data visualization toolset on top of A-Frame.
The editor's user interface is implemented with Preact.js, a JavaScript library that provides a thin Virtual DOM abstraction on top of the Document Object Model (DOM), allowing component-based reactive programming for web user interfaces.
The system follows a Model-View-ViewModel (MVVM) design pattern and any modifications made to the scene through the editor UI will update the VR scene's source data, triggering an update and a re-render for the VR scene.

The backend of VRStory is built with Express.js and Docker to store and serve the media files for the frontend and provides four ML services to facilitate the content generation, namely text summarization, text-to-image, voice generation, and style transfer. 
To concentrate on our primary research objective, VRStory utilizes mature ML models through OpenAI's APIs, rather than investing resources in developing and integrating ML techniques internally.
By leveraging OpenAI's models including \textit{gpt-3.5-turbo} for text generation, \textit{dall-e-2} for image generation, and \textit{tts-1} for text-to-speech, the system enables users to condense lengthy texts into concise summaries, create images from text prompts and produce audio narrations from their text inputs.
Additionally, users can utilize the style transfer function by uploading an image and a style image, which is implemented using Huang \etal's model \cite{huang2017arbitrary}.

\section{User Study: Exploring VR Presentation with VRStory} \label{sec:user-study}
As the final stage of our research (Figure~\ref{fig:flow}), we conducted a user study involving 12 participants to investigate the potential of VR presentations through hands-on working sessions with VRStory. 
The objective of this user study was to validate our findings in the formative study and further gather users' opinions and insights into VR presentations.

\subsection{Study Design and Setup}
This study consisted of two parts, each with six participants. 
In Part 1, participants were guided to utilize VRStory's core features through optimal workflow to create a VR presentation based on predefined tasks. 
In Part 2, participants built their presentations freely to examine their natural authoring workflow for VR presentations.
The study took place one-on-one in a research lab, where participants used a laptop running Edge browser and VRStory, along with an Oculus Quest 2 headset connected via wired Link.

\subsection{Participants}
We recruited a diverse group of 12 participants (5 females, 7 males) with ages ranging from 20 to 34 ($Md=26.5$) from a local research university through its graduate students' mailing list, targeting those proficient in presentation software and enthusiastic about exploring VR technology.
In the following, we will refer to these participants as U\#. 
Specifically, U1, U3, U4, and U5 have experience in creating VR applications; and U2, U6, and U8 possess experience using VR before this study. 
Participants were compensated with \$20 for their time.

\subsection{Procedure}
The user study lasted about 50 minutes and consisted of three phases, where Parts 1 and 2 only differentiated in Phase~1. Specifically, Phase~1 (30 minutes) included scenario-based guided tasks for Part 1 or authoring from scratch for Part 2. Phase~2 (5 minutes) was a free exploration with a premade VR presentation. Phase~3 (15 minutes) was a semi-structured interview. 
For Phases 1 \& 2, a think-aloud protocol was employed to obtain participants' feedback and thoughts on the fly.

Prior to the study, a concise introduction to VR presentations and VRStory was offered. 
In Phase 1, participants in Part 1 undertook a set of tasks to create a VR presentation by following similar steps outlined in \autoref{sec:system-walkthrough}. 
As they worked through these tasks, participants had the freedom to explore different choices and adjust parameters according to their interests.
Participants in Part 2 were asked to make a 3-5 minute VR presentation from any topics they chose using VRStory.
In Phase 2, participants freely explored a refined version of the ``Introduction to Impressionism'' VR presentation demonstrating all the interactive elements provided by VRStory. 
Following the exploration, participants completed a questionnaire to provide feedback on their authoring experience with VRStory and their viewing experience of the premade VR presentation.
In Phase 3, a semi-structured interview was conducted to elicit participants' perspectives and reflections on VR presentation and their ideal authoring workflows.



\section{Findings}
In this section, we present our synthesised key findings from both the formative study (\autoref{sec:formative-study}) and the user study (\autoref{sec:user-study}) around our research questions.
We start by summarizing users' perspectives on the common and unique design aspects for VR presentations, and then discuss their opinions towards VR presentations concerning the challenges.  

\subsection{Opinions on Design Aspects}
In \autoref{sec:background-study}, we highlighted three essential design aspects for presentations. Here, we examine and compare these aspects between 2D presentations and VR presentations.

\subsubsection{2D Elements vs. 3D Elements}
While all participants have experience with VR, there was no universal consensus regarding the core elements for VR presentations. 
Initially, participants primarily associated VR presentations with converting traditional 2D content into an immersive virtual environment, mostly focusing on displaying 2D text and images.
They described VR presentations as \pqt{In VR, we can have a big screen,}{P4} \pqt{another display or another channel of information,}{U6} and \pqt{using VR to give a presentation.}{U11} 
Following the user studies and getting hands-on experience with VR presentations, participants' viewpoints evolved, demonstrating stronger interest in 3D elements and immersive features.
For instance, U9 remarked, \qt{It (VR presentation) has many dimensions we can use and it's more interactive with the listener and more connected with the information we share with them.}

All participants agreed that VR can enhance presentations, even when limited to 2D elements. 
P1 who is an experienced researcher in VR, offered a thought-provoking perspective: \qt{A lot of people think VR is only good for 3D, so everything has to be 3D. But I would make an analogy to when you go and watch a 3D movie in the cinema. There's only $10 \%$ maximum where they actually really work with depth perception. And maybe it's the same for VR presentation.}
Some participants still urged that there should be more exploration of using 3D elements in VR presentation to benefit their professional needs.
P3, as an architecture professor, argued \qt{The prototype can really leverage the idea of virtual space, a space that I can explore more than just a two-dimensional representation.}

\subsubsection{Focused Layout vs. Informative Layout}
Participants who regularly conduct business or team presentations preferred a simple and planar layout that showcases only a few or one element at a time. This preference was emphasized by P6, an experienced associate specializing in IPO services, stating that \qt{Data is the most important thing presented in every presentation. [...] We need to make it into a simple way to help our client to easily understand.}
This perspective was also echoed by U7 and U12, who are seasoned professionals with extensive experience working with data.
Conversely, three university educators preferred a more complex layout incorporating multiple diverse elements for students to explore.
P2 highly praised the immersive nature of VR presentation and noted, \qt{It's a dynamic interface where students are excited to see what's around. [...] I have a textbook that has a lot of pictures I would love for them to experience it in this way.}
This sentiment was echoed by educator U10, who said, \qt{I like the stages actually. [...] If you're in a room, we could have one in the front, one in the back. Just to have more items in a single room.}
Summarized by U4, the presentation layout may largely depend on its purpose: self-paced exploration presentations, lectures requiring clear and quick information transmission, or visually less significant TED Talk-style presentations.

\subsubsection{Controlled Navigation vs. Exploratory Navigation}
There was a notable distinction in participants' views regarding their preference for presentation navigation.
Participants holding roles in business and team communication strongly preferred a straightforward linear structure commonly found in traditional 2D presentations like PowerPoint. 
This format allowed them to maintain complete control over the presentation's pace and sequencing, which they considered crucial for effective communication. 
As an illustration, P4 who is a busy UX designer, directly complained \qt{If you can't control how the audience consumes your content because you gave them the freedom to jump between different slides, sometimes they will be distracted. [...] They may lose some details, and then they will repeat questions again.}

In contrast, P5 who is an architect, favored a spatial and multi-linked navigation for its advantages in allowing audience more agency and flexibility to explore presentations, he remarked: \qt{I think VR is a very nice, actually a perfect way to show the clients how the architecture looks like. [...] They can explore more from the presentation and they can get whatever information they need.}
U6 echoed this idea, commenting \qt{I am like walking around a VR museum. [...] So if there's a presentation or a video by the side, I then can directly learn more details.}
In regard to P7's insightful observation, they suggested that the distinct advantage in spatial navigation largely stems from enhanced ability to make informed decisions, \qt{Exploration may improve the engagement, and eventually, it can also improve decision making.}
The emphasis on ``engagement and exploration'' was also echoed by the three university educators.

\subsection{Opinions around Design Challenges}
In \autoref{sec:design-challenges}, we outlined four key challenges based on our formative study with seven professionals. 
In this subsequent user study, these challenges were reiterated by many participants. 
Here, we present the participants' opinions and feedback regarding VR presentations.

\subsubsection{Mental Model Adaption} \label{sec: mental2}
Non-technical users can struggle to adapt their mental models from 2D presentations to fully utilize the unique capabilities of VR presentations. 
To address this challenge, previous studies utilized a strategy involving starter templates with customizable components  \cite{takala_ruis_2014, zhang_flowmatic_2020, chen_entanglevr_2021, artizzu_defining_2022}.

While other popular tools like Unity and libraries like Three.js can also create VR presentations, participants (U3-U6, U8, and U10-U12) highlighted VRStory's advantage in time efficiency and reduced effort through its predefined templates and modular design. 
These templates serve as a starting point, allowing users to quickly begin creating their VR presentation and customize it to suit their needs. 
The modular design further streamlines the process by breaking it down into manageable components, enabling the efficient development of interactive VR presentations.
U10 commented that \qt{Templates I think can help a lot. Because as I said, making things for VR is very complex, but if you give lots of templates, that makes things easier.}
This approach not only simplified the start for novices but also enabled all the participants to explore and harness the unique capabilities of VR, even for those with experience in VR development.
For instance, U3 remarked \qt{The template definitely made it easier to create a VR presentation. It's like a 3D version of Notion (a popular productivity tool). You can create and insert all those things}.
U4 also appreciated the template approach, saying \qt{I like it's sort of giving you a premade way of making presentations. You don't have anything like that if you are doing similar things with Unity.}
U2, U5, and U8 praised the navigation editor for its visualized templates that enable customizing navigation among stages. U2 appreciated how \qt{it's helpful to have them (stages) not organizing a linear way} while U8 found the scene graph particularly beneficial, saying \qt{Navigation editor makes a lot of sense because not a lot of presentation applications allow to see this kind of view.}

\subsubsection{Accessibility Concerns} \label{sec: accessibility2}
Echoing to the accessibility challenges in the formative study, U7, U11, and U12 expressed additional concerns over the potentially limited availability of VR presentations in their professional settings. 
Specifically, they addressed the challenges associated with adopting new technology like VR in traditional industries like banking.
U7 noted, \qt{I think lots of people in my profession, they are pretty old school, so it's really hard to accept something innovative like this quickly,} which highlights the obstacle of getting older or more established professionals to adopt VR presentation.
U12 added that, \qt{I don't know if my bank is willing to pay for the equipment}, emphasizing the financial aspect and potential reluctance from employers to invest in costly hardware and software required for VR presentation.

\subsubsection{Technology Considerations}
\begin{figure*}[tb!]
    \centering
    \includegraphics[width=1\textwidth]{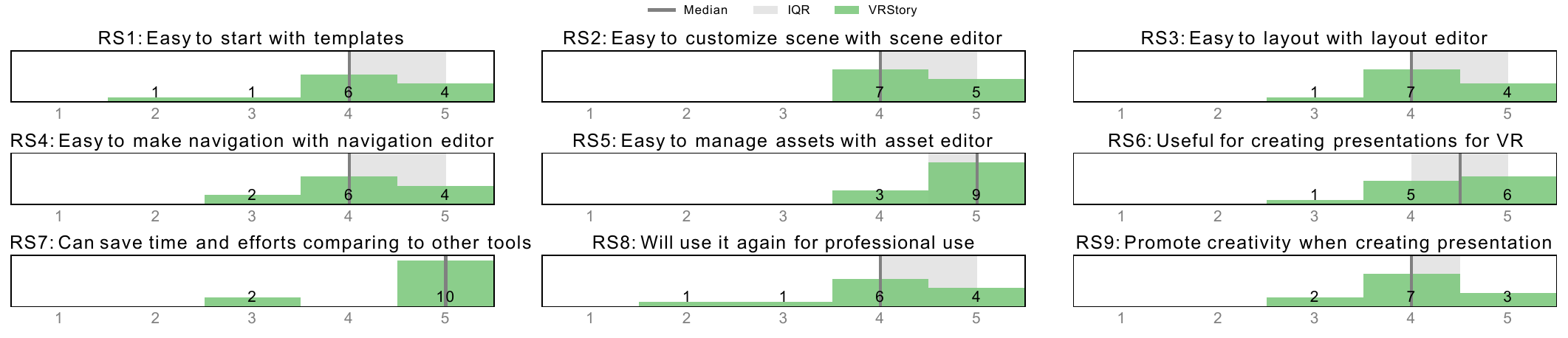}\vspace{-1ex}
    \caption{
   Questionnaire Results: RS1 to RS5 for users' general feedback for each editing modules in VRStory; RS6 to RS9 for users' general experience and attitudes towards VRStory (5-point Likert scale, 1=strongly disagree, 5=strongly agree).
    }
    \label{fig:result}
\end{figure*}

VRStory has demonstrated as a user-friendly non-coding authoring tool, effectively addressing the technology challenges.
As shown in Figure \ref{fig:result}, participants in the user study rated various aspects of VRStory using a 5-point Likert scale. 
Generally, all 12 participants offered positive feedback on the core features (RS1-5).
Moreover, U1, U2, U6, U8, and U9 found VRStory easy and intuitive to use, particularly for non-technical users as U6 commented that \qt{The tool makes it more accessible for end users. It's more like a pipeline. [...] Definitely, it's helpful and it just makes everything easier.}
Participants held a highly positive view ($Md=4.5$) regarding VRStory's usefulness in creating VR presentations (RS6) and its ability to save time and effort in the authoring process (RS7) compared to other existent tools. 
Participants also were willing to use VRStory for their future professional work (RS8) and believed that it promoted their creativity in creating VR presentations (RS9).
These results all suggested the crucial roles of appropriate tools in overcoming technological obstacles and enhancing user engagement with VR presentations.

While the participants generally praised the design and utility of VRStory, 
participants' familiarity with 2D presentations continued to hinder their discovery and utilization of VR features when using VRStory.
Among all the participants, we only observed U8 to demonstrate an initiative to delve into utilizing 3D models and immersive environments on his own accord without any prompting or encouragement from us.
All other participants primarily focused on interacting with 2D images and 2D text, overlooking other available VR features offered by VRStory.
U1, U7, U11, and U12 suggested that VRStory should enhance its UI's readability and familiarity by aligning it more closely with the conventions of traditional 2D presentation tools. 
U11 also expressed a preference for starting with an empty stage without any templates, which aligned with their usual approach and habits when creating 2D presentations.

\subsubsection{Contents Generation} \label{sec: ai-concern2}
To address the challenges of creating high-quality content for VR presentations, VRStory was built with various content generation methods. 
These techniques received widespread praise from the participants, as U7 remarked \qt{I think the most useful feature in this tool is generating the pictures and summarizing the voice. [...] because VR presentation is somewhat very innovative that nobody has really used massively.}

While participants acknowledged the advantages of ML methods for content generation, they voiced concerns over legal implications and the quality of generated materials, which echoed to the formative study's findings. 
All participants preferred maintaining final control over the generated content and viewed ML agents as creative collaborators instead of dominant authorities.
In numerous situations, the generated content failed to fulfill users' specific requirements or accurately convey their design intentions, requiring subsequent editing by the user: \pqt{the generation currently is not generating actually the thing I really want.}{U1} 
U3, and U11 highlighted copyright worries when utilizing ML-generated contents, expressing concerns that the additional effort required to verify the content may outweigh the benefits of using ML methods, as U11 suggesting \qt{Probably a search box could be more practical than an AI generation, so they can search for images that they're exactly looking for.}



\section{Discussions}
In this section, we begin by sharing our learned design considerations for upcoming technologies aimed at facilitating VR presentations based on our findings. 
Following this, we will discuss the limitations of our research and suggest possible future work.

\subsection{Design Considerations}
In our study, we uncovered four design challenges for VR presentations that were consistently encountered among all participants. 
To address these challenges, we built VRStroy to explore potential solution strategies and collected user feedback on these approaches. 
We now present three derived design considerations corresponding to these challenges for future VR presentation design and development.

\subsubsection{Expose and Promote VR Features}
In our study (\autoref{sec: mental1} and \autoref{sec: mental2}), it was observed that many participants persist in using established mental models for 2D presentations and do not proactively delve into VR specific features, even when presented with user-friendly tools. 
This could be due to personal or professional preferences aimed at maintaining concise and clear communication for improved efficiency.
However, to fully exploit VR's capabilities and leverage its unique strengths in VR presentations, future designs could actively highlight the provided VR features and motivate users to experiment with and incorporate them into their work. 
Previous research \cite{nebeling_trouble_2018, kraus_elements_2022} has highlighted the difficulties faced by users when selecting XR authoring tools, ranging from low-fidelity to high-fidelity, due to the vast landscape of options available.
By considering users' technical skills and requirements, it is essential to proactively expose and promote VR features relevant to specific tasks, which can better assist and guide users in selecting the most suitable tools from different fidelity levels.
Additionally, using predefined templates and customizable modular components may also be an effective approach, as explored in both previous \cite{takala_ruis_2014, zhang_flowmatic_2020, artizzu_defining_2022} and our research.
This approach could help users become more comfortable with adopting VR or other advanced features that may initially seem unfamiliar to them.

\subsubsection{Prioritize Accessibility over Immersiveness}
While promoting VR features is crucial for an optimal immersive experience for VR presentations, as frequently advocated by many study participants (\autoref{sec: accessibility1} and \autoref{sec: accessibility2}) , future designs should place greater focus on two types of accessibility over advanced features.
First, VR presentations should be designed to accommodate both 2D displays and VR modes simultaneously by default, which will allow for wider device compatibility and caters to diverse viewing preferences.
While authoring VR contents directly within VR has been widely explored and demonstrates strong potential for efficiency and user preference \cite{zhang_flowmatic_2020, xia_spacetime_2018, nebeling_xrdirector_2020, farias_immersive_2022, coelho_authoring_2022}, as Biener et al.'s week-long study highlighted \cite{biener_quantifying_2022}, current VR devices have limitations such as physical discomfort and mental frustration that hinder users from working in VR for extended periods. 
Therefore, enabling users to create and consume VR content through 2D displays is still crucial for future VR presentation tools to better suit different working preferences.
Secondly, though VR is ideal for integrating spatial interactions and body movements, VR presentation should prioritize compatibility with a stationary or seated viewing experience with minimal physical movements.
Additionally, VR interactions that rely on gaze, hands, or controllers should also accommodate traditional input methods such as mice, keyboards, and touchscreens, which allows users without VR devices to easily interact with VR presentation and access the content.
By adopting this approach, viewers can appreciate the core content of the presentation regardless of their accessibility needs or physical limitations, making VR presentations more accessible to a broader audience.

\subsubsection{Provide Transparency and Control over Generated Contents}
Despite the growing body of research on AI in XR, Hirzle et al.'s recent survey \cite{hirzle_when_2023} highlights that most current AI-XR research focuses on optimizing system efficiency and performance, calling for more studies on usability and user experience aspects in AI-supported XR applications.
The rapid advancement of AI-empowered content generation offers a promising solution to overcome challenges in sourcing and creating high-quality multimedia elements required for VR presentations, which can significantly enhance the authoring experience and make high-quality VR presentations more accessible even for non-technical users.
However, concerns about the suitability of generated contents for professional settings can also arise among users. 
As many study participants noted (\autoref{sec: ai-concern1} and \autoref{sec: ai-concern2}), these concerns go beyond ensuring the quality of generated contents to meet professional requirements for direct use in presentations. 
They also involve potential legal and copyright issues with using these contents and strong ethical concerns about fabricating manipulated multimedia for history and social events.
To address these concerns, future design should be transparent about the usage of content generation and other ML methods while giving users ultimate control over these tools and their outputs.
This way, users can have more confidence in the quality and appropriateness of their created content, so they can be more comfortable to integrate and use ML tools within their professional settings.

\subsection{Limitations and Future Work}
In our research, we followed a four-step approach to investigate the potential of VR presentations and explore different strategies to support non-technical users to make VR presentations with ease. 
While VRStory exhibited promising results, it is crucial to acknowledge its limitations and those inherent in our study. Here, we discuss these limitations and suggest future research directions to advance VR presentations and their authoring experience.

\subsubsection{Investigate More Completed Process for VR Presentation.}
Our study primarily investigates the potential of VR presentations during the creation phase of an entire presentation workflow. 
It's important to acknowledge that other crucial phases, such as distribution and live delivery, have the potential for further exploration to better understand VR presentations' capabilities.
Our research observed a strong interest among participants in experiencing collaborative, multiplayer VR presentation sharing, which could facilitate group activities like joint presentations or asynchronous collaboration in education and business settings. 
Future work can delve into the design and development of these multi-user experiences, allowing multiple individuals to interact with shared objects and environments within a VR presentation.
Another promising direction is examining the distribution of VR presentations, particularly offline distribution through local executable files, which has significant potential in educational settings by empowering students to access and independently explore presentations without internet connectivity. 
Future work could entail developing such a tool and assessing its impact on education, training, and other related domains.

\subsubsection{Explore More Advanced and VR Related ML Models.}
ML-based content generation can significantly improve authoring experiences and unlock creative possibilities in VR presentations for non-technical users. 
However, our study might have utilized outdated ML models, resulting in suboptimal performance compared to the current state-of-the-art (SOTA) approaches.
Future work could explore and fine-tune SOTA models like OpenAI's GPT-4 or Stability AI's SDXL, which can enhance generated content quality to better align with users' design goals, expanding the capabilities of VR presentations for easier and more efficient high-quality content creation.
Another limitation in our study involves primarily focusing on ML models generating basic assets, such as images and audio files. 
Future work could explore more advanced ML techniques like Neural Radiance Fields (NeRF) to generate more advanced assets, including panorama images, 3D environments, and 3D models.

\subsubsection{Extend to XR and More Scenarios.}
While our research primarily concentrates on VR presentations, the emergence of commercial Extended Reality (XR) devices, such as Oculus Quest 3 and Apple Vision Pro, is blurring the distinction between VR and augmented reality (AR) in this context. 
This transition from VR to XR presentations opens up significant opportunities across various scenarios like museums, education, and group training, calling for future research to explore its application scenarios and social implications.
Although transitioning a VR application to an XR one may be relatively simple from a development perspective, it can become significantly challenging for non-technical users to grasp and create XR presentations from the authoring side and user standpoint, primarily due to their developed mental models around 2D presentations. 
It can be crucial and promising to investigate methods to support non-technical users in understanding and creating XR presentations tailored to their domain tasks.
This could entail developing user-friendly and intuitive authoring tools and interaction techniques that can guide and scaffold users throughout the authoring process, making the transition from 2D presentations more accessible and seamless.

\section{Conclusion}
Anticipating the growing interests in using VR to deliver presentations, we present a four-step investigation into the potential of VR presentations. 
Our research involved a background study on popular presentation tools, interviews with seven professionals, and a user study with 12 participants using a prototype authoring tool.
We identified and explored five design aspects and four design challenges for VR presentations.
Our findings reveal that VR has the potential to enhance presentation experience and effectiveness for a broad audience, when VR presentations are effectively designed to help users shift their mental models from traditional 2D presentations and address accessibility concerns.
Our research shed light upon the challenges and opinions users engaging in the growing trend of using VR to present information, promoting further exploration and research within this related field.

\begin{acks}
This research is supported in part by the NSERC Discovery Grant (RGPIN-2020-03966), Meta gift fund, and Cisco gift fund.
\end{acks}

\bibliographystyle{ACM-Reference-Format}
\bibliography{00-base}

\appendix

\end{document}